\documentclass[aps,prc,10pt,showpacs,showkeys,twocolumn,superscriptaddress,groupedaddress]{revtex4-1}
\usepackage[english]{babel}
\usepackage{graphicx,psfrag}
\usepackage{amsmath}
\usepackage{amssymb}
\usepackage{amscd}
\usepackage{eucal}
\usepackage{float}
\usepackage{color}
\usepackage{bm}
\usepackage{braket}
\usepackage{dsfont}
\usepackage{hyperref}
\usepackage{placeins}
\usepackage{isotope}
\usepackage{subfigure}
\usepackage[export]{adjustbox}

\begin{document}
\title{Microscopic investigation of the $^8$Li($n, \gamma$)$^9$Li reaction}

\author{Callum McCracken}
\email{ccmccrac@uwaterloo.ca}
\affiliation{TRIUMF, 4004 Wesbrook Mall, Vancouver, British Columbia V6T 2A3, Canada}
\affiliation{University of Waterloo, 200 University Ave, Waterloo, Ontario N2L 3G1, Canada}

\author{Petr Navr\'atil}
\email{navratil@triumf.ca}
\affiliation{TRIUMF, 4004 Wesbrook Mall, Vancouver, British Columbia V6T 2A3, Canada}

\author{Anna McCoy}
\email{amccoy@triumf.ca}
\affiliation{TRIUMF, 4004 Wesbrook Mall, Vancouver, British Columbia V6T 2A3, Canada}

\author{Sofia Quaglioni}
\email{quaglioni1@llnl.gov}
\affiliation{Lawrence Livermore National Laboratory, P.O. Box 808, L-414, Livermore, California 94551, USA}

\author{Guillaume Hupin}
\email{hupin@ipno.in2p3.fr}
\affiliation{Universit\'e Paris-Saclay, CNRS/IN2P3, IJCLab, 91405 Orsay, France}

\date{\today}

\begin{abstract}

\noindent
{\bf Background:} The $^8$Li($n,\gamma$)$^9$Li reaction plays an important role in several astrophysics scenarios. It cannot be measured directly and indirect experiments have so far provided only cross section limits. Theoretical predictions differ by an order of magnitude.

\noindent
{\bf Purpose:} In this work we study the properties of $^9$Li bound states and low-lying resonances and calculate the $^8$Li($n,\gamma$)$^9$Li cross section within the no-core shell model with continuum (NCSMC) with chiral nucleon-nucleon and three-nucleon interactions as input.

\noindent
{\bf Methods:} The NCSMC is an {\it ab initio} method applicable to light nuclei that provides a unified description of bound and scattering states well suited to calculate low-energy nuclear scattering and reactions. For the capture cross section calculation, we adjust calculated thresholds to experimental values. 

\noindent
{\bf Results:} Our calculations reproduce the experimentally known bound states as well as the lowest $5/2^-$ resonance of $^9$Li. We predict a $3/2^-$ spin-parity assignment for the resonance observed at 5.38 MeV. In addition to the a very narrow $7/2^-$ resonance corresponding presumably to the experimental 6.43 MeV state, we find several other broad low-lying resonances.

\noindent
{\bf Conclusions:} Our calculated $^8$Li($n,\gamma$)$^9$Li cross section is within the limits derived from the 1998 National Superconducting Cyclotron Laboratory Coulomb-dissociation experiment [Phys. Rev. C {\bf 57}, 959 (1998)]. However, it is higher than cross sections obtained in recent phenomenological studies. It is dominated by a direct E1 capture to the ground state with a resonant contribution at $\sim0.2$ MeV due to E2/M1 radiation enhanced by the $5/2^-$ resonance.

\end{abstract}


\maketitle

\section{Introduction}
\label{sec_intro}

In neutron rich astrophysical environments, reactions involving the short-lived $^8$Li nucleus 
may contribute to the synthesis of heavier nuclei by bridging the stability gap of mass $A=8$ elements. In particular, the $^8$Li($n, \gamma$)$^9$Li capture reaction plays an important role in inhomogeneous big bang 
nucleosynthesis and in the r-process. 
There, it competes with the $^8$Li($\alpha, n$)$^{11}$B reaction and the $^8$Li beta decay, affecting the reaction path to $A{>}8$ isotopes and also the abundances of Li, Be, B, and C. The relevant reaction chains are $^7$Li($n,\gamma$)$^8$Li($\alpha,n$)$^{11}$B($n,\gamma$)$^{12}$B($\beta^+$)$^{12}$C and $^7$Li($n,\gamma$)$^8$Li($n,\gamma$)$^9$Li($\alpha,n$)$^{12}$B($\beta^+$)$^{12}$C~\cite{PhysRevLett.68.1283,GU199531,Malaney1988,Rauscher1994}. In addition, the reaction chain with two-neutron captures $^4$He($2n,\gamma$)$^6$He($2n,\gamma$)$^8$He($\beta^-$)$^8$Li($n,\gamma$)$^9$Li($\beta^-$)$^9$Be, of which the $^8$Li($n, \gamma$)$^9$Li is also a component, has been considered as an alternative 
to the triple-alpha process in overcoming the $A=8$ mass gap in the r-process 
for supernovae of type II~\cite{PhysRevC.52.2231,Efros1996}. 

As the $^8$Li half-life is 840 ms and a neutron target is not available, the $^8$Li($n, \gamma$)$^9$Li reaction cannot be measured directly. There have been several attempts to determine its cross section by indirect methods. Using a radioactive beam of $^9$Li and the Coulomb-dissociation method with U and Pb targets, only upper limits on the $^8$Li($n, \gamma$)$^9$Li cross section were determined as it was not possible to estimate the nuclear contribution to the dissociation~\cite{Zecher1998}. A follow-up Coulomb-dissociation experiment using a Pb target reported a null result and consequently a very low limit on the capture cross section~\cite{Kobayashi2003}.

In Ref.~\cite{Li2005}, the direct $^8$Li($n,\gamma$)$^9$Li$_{\rm g.s.}$ capture cross section was 
computed in the framework of the potential model by deducing the single particle spectroscopic factor for the ground state of $^9$Li from a measurement of the angular distribution of the $^8$Li($d, p$)$^9$Li$_{\rm g.s.}$ transfer reaction at $E_{\rm c.m.}{=}7.8$ MeV. 
The obtained reaction rate was lower than the limit from Ref.~\cite{Zecher1998} but significantly higher than the limit from Ref.~\cite{Kobayashi2003}. A similar extraction, but with the spectroscopic factor obtained from 
the angular distribution of the $^9$Be($^8$Li,$^9$Li)$^8$Be transfer reaction 
measured with a 27 MeV $^8$Li radioactive nuclear beam, was reported in Ref.~\cite{Guimaraes2007}. 
The obtained reaction rate was comparable to that from Ref.~\cite{Li2005}. 

There were several other studies focused on the structure of $^9$Li. Notably, the $^2$H($^8$Li,$p$)$^9$Li reaction with 76 MeV radioactive $^8$Li beam was studied with the goal to obtain single-neutron spectroscopic factors for states in $^9$Li~\cite{Wuosmaa2005}. Spectroscopic factors for the $^9$Li ground state have also been investigated through the $d$($^9$Li, $t$)$^8$Li one-neutron transfer reaction at E/A = 1.68 MeV~\cite{KANUNGO200826}. The first excited state of $^9$Li was studied by the inelastic scattering of $^9$Li from deuterons~\cite{ALFALOU2013224}. A very recent experiment investigated the structure of $^9$C, the mirror of $^9$Li, using proton resonant scattering~\cite{Hooker2019}. 

The $^8$Li($n,\gamma$)$^9$Li cross section and its reaction rate have also been the focus of several theoretical investigations, based on various approaches. In Refs.~\cite{Malaney1989,Rauscher1994}, 
the reaction rate was estimated based on the existing information for other nuclei. 
Calculations combining the shell model and the potential model were reported in Refs.~\cite{MAO1991568,Ma2012}. In both these studies, multi-major shell model spaces were used. Their predicted reaction rate, however, differed significantly, with the former reporting the rate about five times higher than the latter. The potential model was also applied to a simultaneous study of the $^8$Li($n,\gamma$)$^9$Li reaction and its mirror, 
$^8$B($p,\gamma$)$^9$C~\cite{Mohr2003} using consistent potential parameters. This study revealed a sensitivity of the $^8$Li($n,\gamma$)$^9$Li cross section to the strengths of the potential. 
In Ref.~\cite{Descouvemont1993}, the neutron capture on $^8$Li was investigated by means of the microscopic cluster model again with a simultaneous study of the $^8$B($p,\gamma$)$^9$C mirror reaction. In this approach, the Pauli principle is exactly taken into account. The obtained reaction rate was higher than that of Ref.~\cite{Ma2012}. However, contrary to present experimental knowledge, the $5/2^-$ state was predicted as bound. The Coulomb dissociation of $^9$Li on heavy targets was calculated using a potential model for $^9$Li in Refs.~\cite{Bertulani_1999,Banerjee2008} and the principle of detailed balance was then used to obtain the $^8$Li($n,\gamma$)$^9$Li reaction rate with results reported in the two studies differing by about 50\%. More recently, this reaction was investigated within the framework of the modified potential cluster model with the state classification of nucleons according to the Young tableaux~\cite{Dubovichenko_2016}. Multiple potential  parametrizations were explored with calculated cross sections within the upper limits obtained in Ref.~\cite{Zecher1998}. Overall, predictions of the reaction rate by various theoretical approaches span more than an order of magnitude.

In this work, we report the first {\it ab initio} calculation of the $^8$Li($n,\gamma$)$^9$Li cross section. We apply the no-core shell model with continuum (NCSMC)~\cite{PhysRevLett.110.022505,PhysRevC.87.034326,physcripnavratil} and use chiral nucleon-nucleon (NN) and three-nucleon (3N) interactions as input. In particular, we employ the chiral Hamiltonian from Ref.~\cite{PhysRevC.101.014318} shown to describe well both light and medium mass nuclei. The NCSMC provides a unified description of bound and scattering states and allows us to investigate bound states of $^9$Li as well as its low-lying resonances. 
While in the present evaluation of the capture cross section we adjust calculated thresholds to experimental values, no other experimental information is used unlike in previous studies.

The paper is organized as follows: In Sec.~\ref{sec_formalism} we briefly review the NCSMC formalism. In Sec.~\ref{sec_results}, we present our results for $^8$Li, $^9$Li, and for the capture cross section. Finally, in Sec.~\ref{sec_conclusions} we draw our conclusions.

\section{Theoretical Framework}
\label{sec_formalism}

The starting point of our approach is the microscopic Hamiltonian
\begin{equation}
H=\frac{1}{A}\sum_{i<j=1}^A\frac{(\vec{p}_i-\vec{p}_j)^2}{2m} + \sum_{i<j=1}^A V^{NN}_{ij} + \sum_{i<j<k=1}^A V^{3N}_{ijk} \, ,\label{H}
\end{equation}
which describes nuclei as systems of $A$ non-relativistic point-like nucleons interacting through realistic inter-nucleon interactions. Modern theory of nuclear forces is based on the framework of chiral effective field theory (EFT)~\cite{Weinberg1990,Weinberg1991}. The quantum chromodynamics (QCD) Lagrangian is expanded in powers of 
$Q/\Lambda_{\chi}$, where $Q$ is the characteristic momentum in the nuclear process and $\Lambda_{\chi}\sim 1$ GeV represents the hard scale of the theory. 
Such an expansion allows a systematic improvement of the nuclear interaction and provides a hierarchy of the NN and many-nucleon forces which naturally arises in a consistent scheme~\cite{OrRa94,VanKolck94,EpNo02,Epelbaum06}.

In the present work we adopt the NN+3N chiral interaction applied in Ref.~\cite{PhysRevC.101.014318}, denoted as NN+3N(lnl), consisting of an NN interaction up to the fourth order (N$^3$LO) in the chiral expansion~\cite{Entem2003} and a 3N interaction up to next-to-next-to-leading order (N$^2$LO) using a combination of local and non-local regulators. Even though all the underlying parameters (known as low-energy constants or LECs) are determined in $A{=}2,3,4$ nucleon systems, this interaction provides a very good description of properties of both light and medium mass nuclei~\cite{PhysRevC.101.014318}, including $^{100}$Sn~\cite{Gysbers2019}. The chiral orders of the adopted NN and 3N interactions are not consistent: the former is included up to order N$^3$LO while the latter is at N$^2$LO. While the N$^3$LO 3N contribution has been shown to be rather small~\cite{Machleidt2011}, 
the consistency of the regulator and/or in particular the use of a non-local versus local regulators plays a significant role for medium mass nuclei~\cite{Huther2020}. Even though the NCSMC is formulated in coordinate space, the inclusion of non-local momentum-space NN and 3N interactions is straightforward owing to the use of expansions in harmonic oscillator (HO) basis states, for which the Fourier transformation is trivial~\cite{physcripnavratil}.

A faster convergence of our 
calculations with respect to the many-body basis size is obtained by softening the chiral interaction through the similarity renormalization group (SRG) technique~\cite{Wegner1994,Bogner2007,PhysRevC.77.064003,Bogner201094,Jurgenson2009}. The SRG unitary transformation induces many-body forces, included here up to the three-body level. The four- and higher-body induced terms are small at the $\lambda_{\mathrm{SRG}}{=}2.0$ fm$^{-1}$ resolution scale used in present calculations~\cite{PhysRevC.101.014318}. To verify this, we performed NCSM calculations for several p-shell nuclei varying $\lambda_{\mathrm{SRG}}$ between 1.6 and 2.2 fm$^{-1}$ and found ground-state energy differences of the order of 1\%. Due to the complexity of the NCSMC calculations, results reported in this paper were obtained for a fixed $\lambda_{\mathrm{SRG}}{=}2$ fm$^{-1}$ value.

In the NCSMC~\cite{PhysRevLett.110.022505,PhysRevC.87.034326,physcripnavratil}, the many-body scattering problem is solved by expanding the wave function on continuous microscopic-cluster states, describing the relative motion between target and projectile nuclei (here $^8$Li and the neutron), and discrete square-integrable states, describing the static composite nuclear system (here $^9$Li). The idea behind this generalized expansion is to augment the microscopic cluster model, which enables the correct treatment of the wave function in the asymptotic region, with short-range many-body correlations that are present at small separations, mimicking various deformation effects that might take place during the reaction process. The NCSMC wave function for $^9$Li is represented as
\begin{align}
\ket{\Psi^{J^\pi T}_{A\texttt{=}9, -\frac{3}{2}}} = &  \sum_\lambda c^{J^\pi T}_\lambda \ket{^{9} {\rm Li} \, \lambda J^\pi T} \nonumber \\
& +\sum_{\nu}\!\! \int \!\! dr \, r^2 
                 \frac{\gamma^{J^\pi T}_{\nu}(r)}{r}
                 {\mathcal{A}}_\nu \ket{\Phi^{J^\pi T}_{\nu r, -\frac{3}{2}}} \,.\label{ncsmc_wf_Li9}
\end{align}
The first term of Eq. (\ref{ncsmc_wf_Li9}) consists of an expansion over square-integrable energy eigenstates of the $^9$Li nucleus indexed by $\lambda$. The second term, corresponding to an expansion over the antisymmetrized channel states in the spirit of the resonating group
method (RGM)~\cite{wildermuth1977unified,TANG1978167,FLIESSBACH198284,langanke1986,PhysRevC.77.044002}, is given by
\begin{align}
\ket{\Phi^{J^\pi T}_{\nu r, -\frac{3}{2}}} = &\Big[ \big( \ket{^{8} {\rm Li} \, \lambda_8 J_8^{\pi_8}T_8} \ket{n \, \tfrac12^{\texttt{+}}\tfrac12} \big)^{(sT)}
Y_\ell(\hat{r}_{8,1}) \Big]^{(J^{\pi}T)}_{-\frac{3}{2}} \nonumber\\ 
&\times\,\frac{\delta(r{-}r_{8,1})}{rr_{8,1}} \, .
\label{eq_Li9_rgm_state}
\end{align}
Here, the index $\nu$ represents all relevant quantum numbers except for those explicitly listed 
on the left-hand side of the equation, and the subscript $-\frac{3}{2}$ is the isospin projection, i.e., $(Z-N)/2$. The coordinate $\vec{r}_{8,1}$ in Eq.(\ref{eq_Li9_rgm_state}) is the separation vector between the $^8$Li target and the neutron.

The translationally invariant eigenstates of the aggregate ($\ket{^{9} {\rm Li} \, \lambda J^\pi T}$) and target 
($\ket{^{8} {\rm Li} \, \lambda_8 J_8^{\pi_8}T_8}$) nuclei are all obtained by means of the no-core shell model (NCSM)~\cite{PhysRevLett.84.5728,PhysRevC.62.054311,BARRETT2013131} using a basis of many-body HO wave functions with the same frequency, $\Omega$, and maximum number of particle excitations $N_{\rm max}$ from the lowest Pauli-allowed many-body configuration. In this work we used the HO frequency of $\hbar \Omega = 20$ MeV found as optimal for $p$-shell nuclei in Ref.~\cite{PhysRevC.101.014318}. 

The discrete expansion coefficients  $c_{\lambda}^{J^{\pi}T}$ and the continuous relative-motion amplitudes $\gamma_{\nu}^{J^{\pi}T} (r)$ are the solution of the generalized
eigenvalue problem derived by representing the Schr\"{o}dinger equation in the model space of the expansions (\ref{ncsmc_wf_Li9})~\cite{physcripnavratil}.
The resulting NCSMC equations are solved by means of the coupled-channel R-matrix method on a Lagrange mesh~\cite{Descouvemont2010,Hesse1998,Hesse2002}.

In general the sum over the index $\nu$ in Eq. (\ref{ncsmc_wf_Li9}) includes all the mass partitions involved in the formation of the composite system $^9$Li, i.e., $^8$Li+$n$, $^7$Li+$n$+$n$, $^6$He+$^3$H etc.  
Here, we limit the present calculations to the $^8$Li+$n$ clusters of Eq.~(\ref{eq_Li9_rgm_state}), which are by far the most relevant for the low-energy $^8$Li($n,\gamma$)$^9$Li capture.  
The channel states for the other mass partitions are energetically closed and their effect is in part 
accounted for by means of the first term in Eq.~(\ref{ncsmc_wf_Li9}). Applications of the NCSMC with three-body clusters and with coupling between different mass partitions can be found, e.g., in Refs.~\cite{Quaglioni_2018} and~\cite{Hupin2019}, respectively.

\section{Results}
\label{sec_results}

\subsection{NCSM calculations for $^{8,9}$Li}

\begin{figure}
\includegraphics[width=0.48\textwidth]{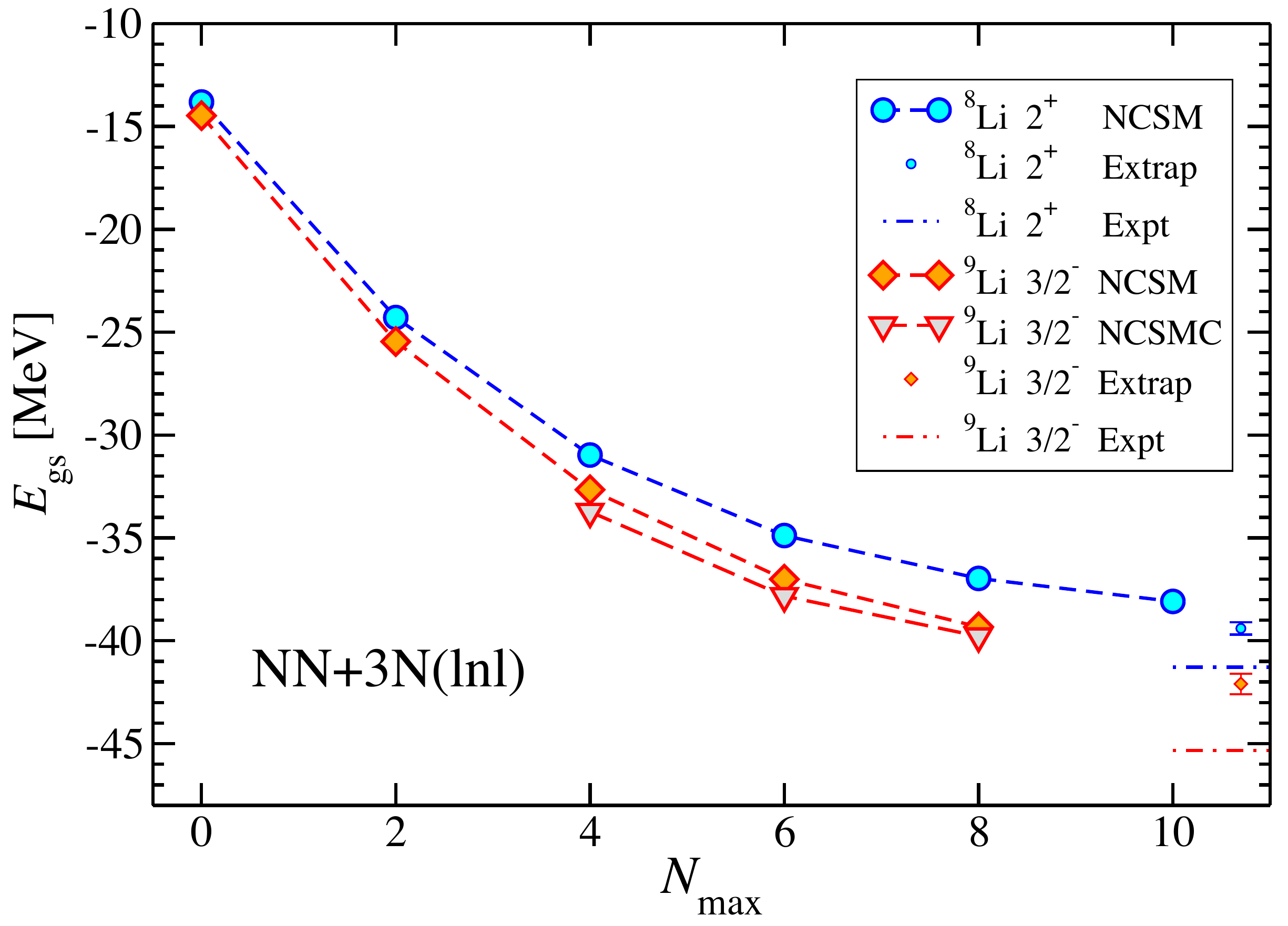}
\caption{\label{fig:Egs} $^8$Li (circles) and $^9$Li (diamonds) ground state energy dependence on the size of the NCSM and for $^9$Li also NCSMC (triangles) basis. Extrapolations to infinite $N_{\rm max}$ with their uncertainties are presented on the right. The experimental values are shown by dashed-dotted lines. The SRG-evolved NN+3N(lnl) chiral interaction~\cite{PhysRevC.101.014318} at the resolution scale of $\lambda_{\mathrm{SRG}} = 2.0$ fm$^{-1}$ and the HO frequency $\hbar\Omega{=}20$ MeV was used. Experimental data are from Ref.~\cite{TILLEY2004155}.}
\end{figure}

The present NCSMC calculations require as input NCSM eigenstates and eigenenergies of $^8$Li and $^9$Li. For $^8$Li, we performed 
calculations up to $N_{\rm max}{=}10$, while for $^9$Li up to $N_{\rm max}{=}8$ and $9$ for the negative- and positive-parity states, respectively. The ground-state energy dependence on the basis size for both isotopes is presented in Fig.~\ref{fig:Egs}. The NCSM extrapolated $^9$Li ground state energy of -42.1(5) MeV for the interaction used here has been reported in Ref.~\cite{PhysRevC.101.014318}. Comparing to the experimental value of -45.34 MeV, the calculation underbinds by a few percent. For $^8$Li we find the ground-state energy -39.4(3) MeV compared to the experimental -41.28 MeV. The theoretical uncertainty is due to the extrapolation to the infinite basis size performed using the exponential function $E (N_{\rm max}) = E_{\infty} + a e^{- b N_{\rm max}}$ and varying the number of $N_{\rm max}$ points.

\begin{figure*}
    \centering
    \includegraphics[width=0.4\textwidth]{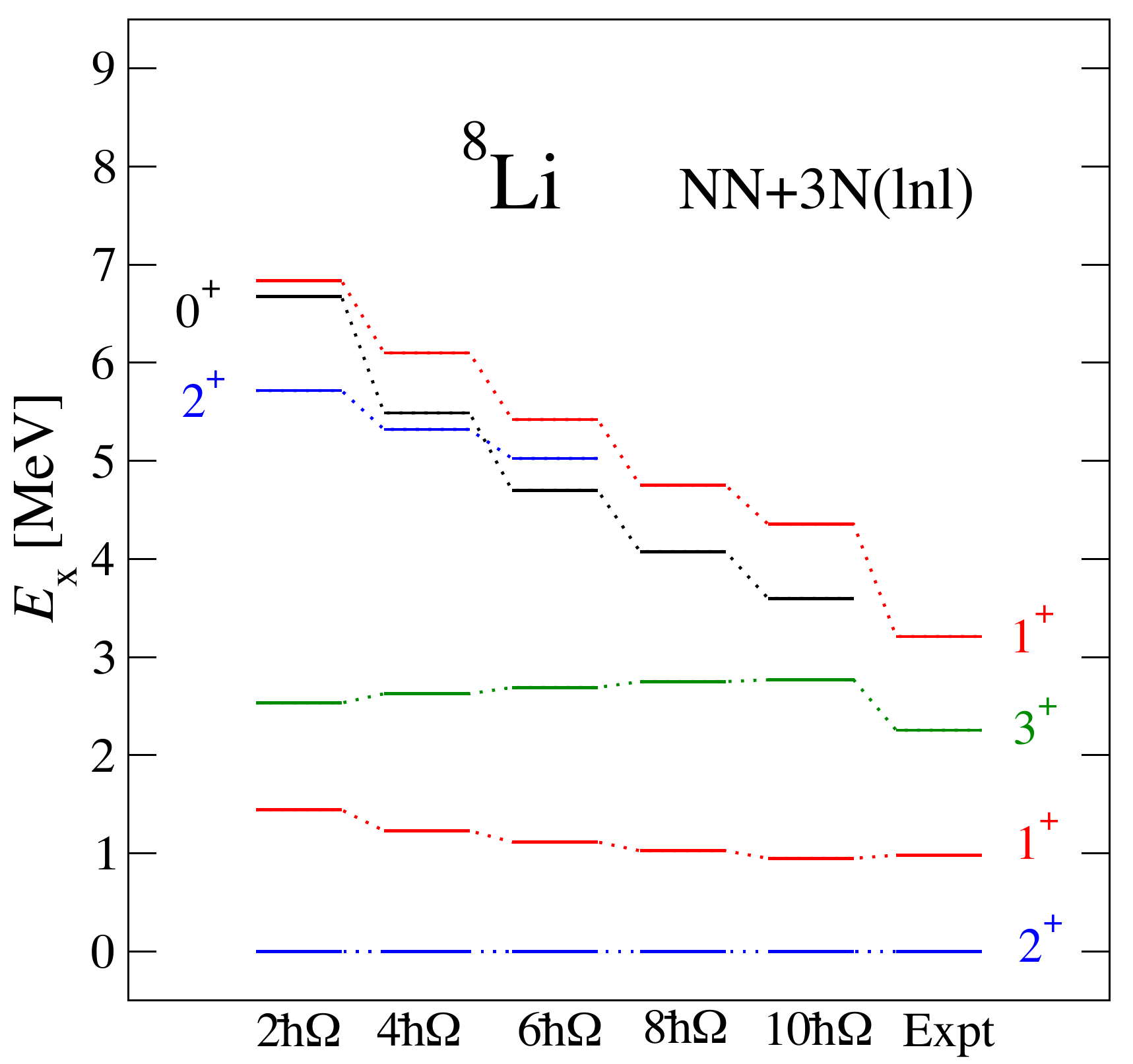}
    \hspace{1cm}
    \includegraphics[width=0.4\textwidth]{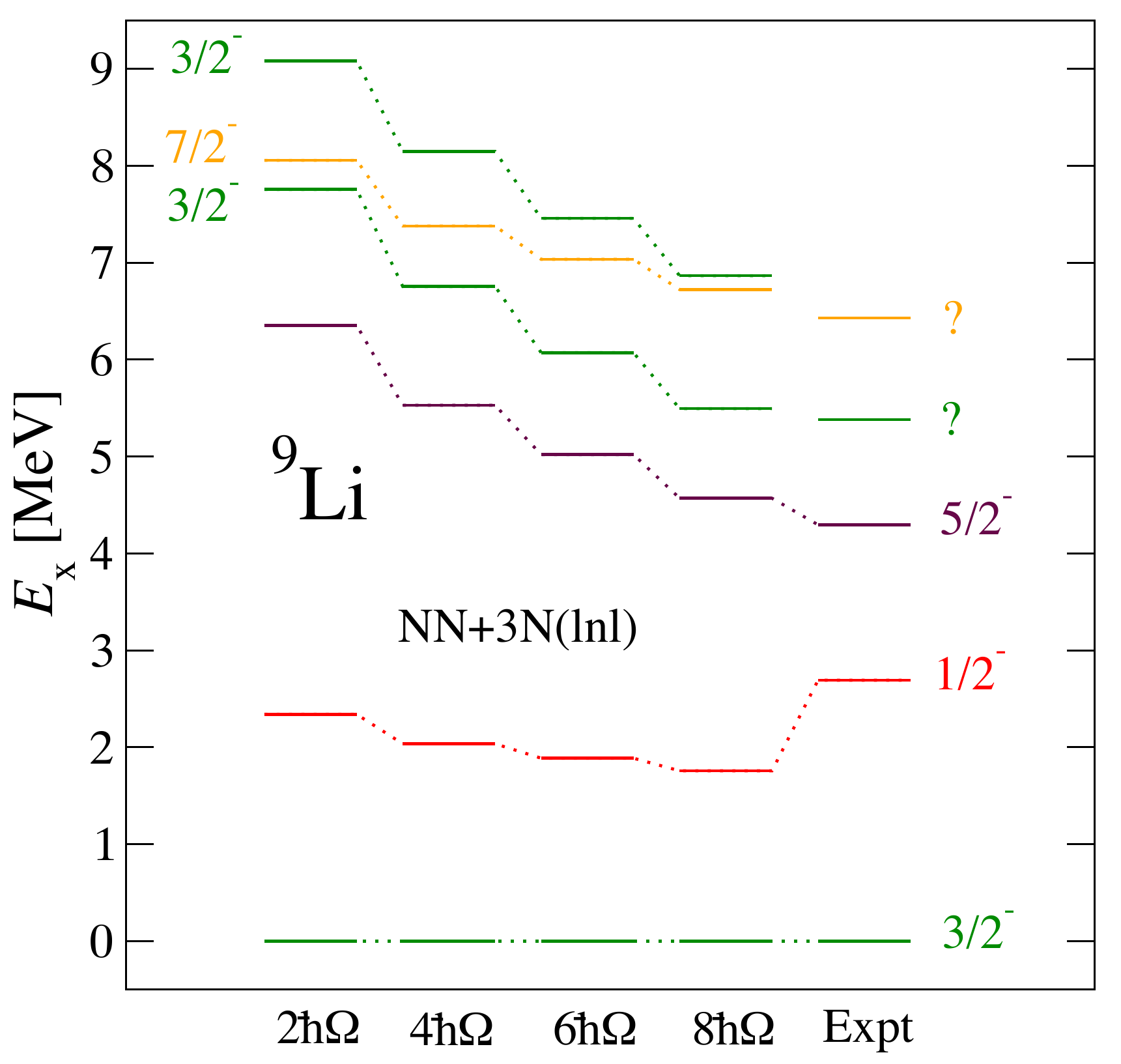}
    \caption{\label{fig:ncsd_spectra} Comparison between the NCSM-calculated and the experimental energy spectra of $^8$Li (left panel) and $^9$Li (right panel). The SRG-evolved NN+3N(lnl) chiral interaction~\cite{PhysRevC.101.014318} at the resolution scale of $\lambda_{\mathrm{SRG}} = 2.0$ fm$^{-1}$. The HO basis frequency was $\hbar\Omega{=}20$ MeV. Experimental data are from Ref.~\cite{TILLEY2004155}.}
\end{figure*}

Excitation energies of $^8$Li low-lying states are shown in the left panel of Fig.~\ref{fig:ncsd_spectra}. The 
convergence of the NCSM approach for the experimentally bound $1^+$ state and the narrow $3^+$ resonance is quite good. The agreement with experiment is quite satisfactory for the $1^+$ state while the excitation energy of the $3^+$ state is overestimated by several hundred keV. The second $1^+$ state is a broad resonance in experiment. In the NCSM calculations, this is reflected by rapid changes of the excitation energy with the size of the model space $N_{\rm max}$. Compared to the known levels, we predict additional states close to the $1^+_2$, most notably a $0^+$ resonance. We note that  both the predicted $0^+$ and the $2^+$ resonances have been previously investigated by studying the $n{+}^7$Li continuum~\cite{PhysRevC.82.034609} working within a predecessor of the NCSMC approach, known as NCSM/RGM.
 Experimental evidence for these resonances in $^8$B, the isospin mirror of $^8$Li, has been reported in Ref.~\cite{PhysRevC.87.054617}.

NCSM results for the low-lying excitation energies of $^9$Li with the interaction used here have been  
reported in Ref.~\cite{PhysRevC.101.014318}. For completeness, we present the negative-parity level energies in the right panel of Fig.~\ref{fig:ncsd_spectra}. The convergence of the experimentally bound $1/2^-$ state is satisfactory, though the experimental $1/2^- - 3/2^-$ splitting is underestimated in the calculation. We find the $5/2^-_1$ state quite close to the experimentally established $5/2^-$ resonance. In addition, we predict a $3/2^-$ and a $7/2^-$ level that might correspond to the experimentally observed resonances at 5.38 MeV and 6.43 MeV with undetermined spins and parities. 

Calculated ground state properties of the two isotopes and the $M1$ transition rate between their bound states are summarized in Table~\ref{tabNCSM8Li9Li}. Only one-body transition operators were used. Overall agreement with experiment is quite reasonable. The magnetic dipole moment discrepancies could be attributed to the missing two-body currents~\cite{Pastore:2015dho} while the underestimation of the quadrupole moments is most likely due to the limited basis size. The calculations should also be in general improved by the SRG evolution of the transition operators~\cite{Schuster2014,Schuster2015,Gysbers2019}.  

\begin{table}
\begin{center}
\begin{ruledtabular}
\begin{tabular}{lcccc}
     &  $E_{\rm g.s.}$ [MeV] & $Q$ [$e$ fm$^2$] & $\mu$ [$\mu_{\rm N}$] & $B(M1)$ [$\mu_{\rm N}^2$] \\
\hline
\multicolumn{5}{}{}$^8$Li \\
NCSM & -39.4(3) &  +2.95(15) & +1.48  & 4.164 \\
Expt & -41.28   &  +3.14(2)  & +1.654 & 5.0(16) \\ 
\hline
\multicolumn{5}{}{}$^9$Li \\
NCSM &  -42.1(5) &  -2.5(2)  & +2.91  & 3.23 \\
Expt &  -45.34   &  -3.06(2) & +3.437 &  N/A \\
\end{tabular}
\end{ruledtabular}
\caption{$^{8,9}$Li ground state energies, quadrupole and magnetic moments, and the $M1$ transition rate between their bound states. In particular, $B(M1;1^+{\rightarrow}2^+)$ and $B(M1;1/2^-{\rightarrow}3/2^-)$ for $^8$Li and $^9$Li, respectively, is shown. NCSM calculations have been performed with the NN+3N(lnl) chiral interaction. Experimental results are from Refs.~\cite{TILLEY2004155,PhysRevC.72.044309}.}
\label{tabNCSM8Li9Li}
\end{center}
\end{table}

For the microscopic cluster component of the NCSMC expansion, Eq.~(\ref{eq_Li9_rgm_state}), we used two NCSM eigenstates corresponding to the two $^8$Li bound states, the $2^+$ ground state and the $1^+$ excited state. In principle, we could have included also the experimentally narrow $3^+$ state. However, since our focus is on the low-energy $^8$Li($n, \gamma$)$^9$Li radiative capture, the impact of the $3^+$ state is expected to be negligible while the technical complexity of the calculations would increase substantially. As for the composite $^9$Li states entering the expansion (\ref{ncsmc_wf_Li9}), we used the eight lowest negative-parity and six lowest positive-parity NCSM eigenstates of $^9$Li with total angular momentum  $J \in \{ 1/2 , 3/2 , 5/2 , 7/2 \}$ and isospin $T{=}3/2$.

\subsection{NCSMC results for $^9$Li}

We performed NCSMC calculations for $^9$Li for $N_{\rm max}{=}4,6,8$ basis spaces. The $^9$Li NCSM positive-parity states entering the expansion~(\ref{ncsmc_wf_Li9}) were obtained in $N_{\rm max}{+}1$ spaces, i.e., up to $N_{\rm max}{=}9$. We found two bound states, the $3/2^-$ ground state and the $1/2^-$ excited state, in agreement with experiment. The NCSMC ground-state energies are shown in Fig.~\ref{fig:Egs} and the separation energies with respect to the $^8$Li${+}n$ threshold for both the $3/2^-$ and $1/2^-$ states are given in Table~\ref{tab:sepen}. NCSMC calculations increase the binding energies compared to the NCSM results at any fixed $N_{\rm max}$ due to the inclusion of the cluster basis component. The separation energies are quite stable with varying $N_{\rm max}$. The calculated $1/2^-$ separation energy is quite close to the experimental one while the ground state separation energy is underestimated by about 1.2 MeV. This could be due to a weaker spin-orbit strength and/or missing strength in the $T{=}3/2$ part of the 3N interaction. 
\begin{table}
\begin{center}
\begin{ruledtabular}
\begin{tabular}{ccccc}
 $J^\pi\; T$    & $N_{\rm max}{=}4$ & $N_{\rm max}{=}6$ & $N_{\rm max}{=}8$ & Expt \\
 $1/2^-\; 3/2$  & -0.98  &  -1.09  &  -1.14  &  -1.37 \\  
 $3/2^-\; 3/2$  & -2.76  &  -2.94  &  -2.81  &  -4.06 \\
\end{tabular}
\end{ruledtabular}
\caption{$^{9}$Li bound-state energies, in MeV, with respect to the $^8$Li${+}n$ threshold. NCSMC calculations have been performed with the NN+3N(lnl) chiral interaction~\cite{PhysRevC.101.014318} at the resolution scale of $\lambda_{\mathrm{SRG}} = 2.0$ fm$^{-1}$. The HO basis frequency was $\hbar\Omega{=}20$ MeV. Experimental data are from Ref.~\cite{TILLEY2004155}.}
\label{tab:sepen}
\end{center}
\end{table}

\begin{figure}
\includegraphics[width=0.46\textwidth]{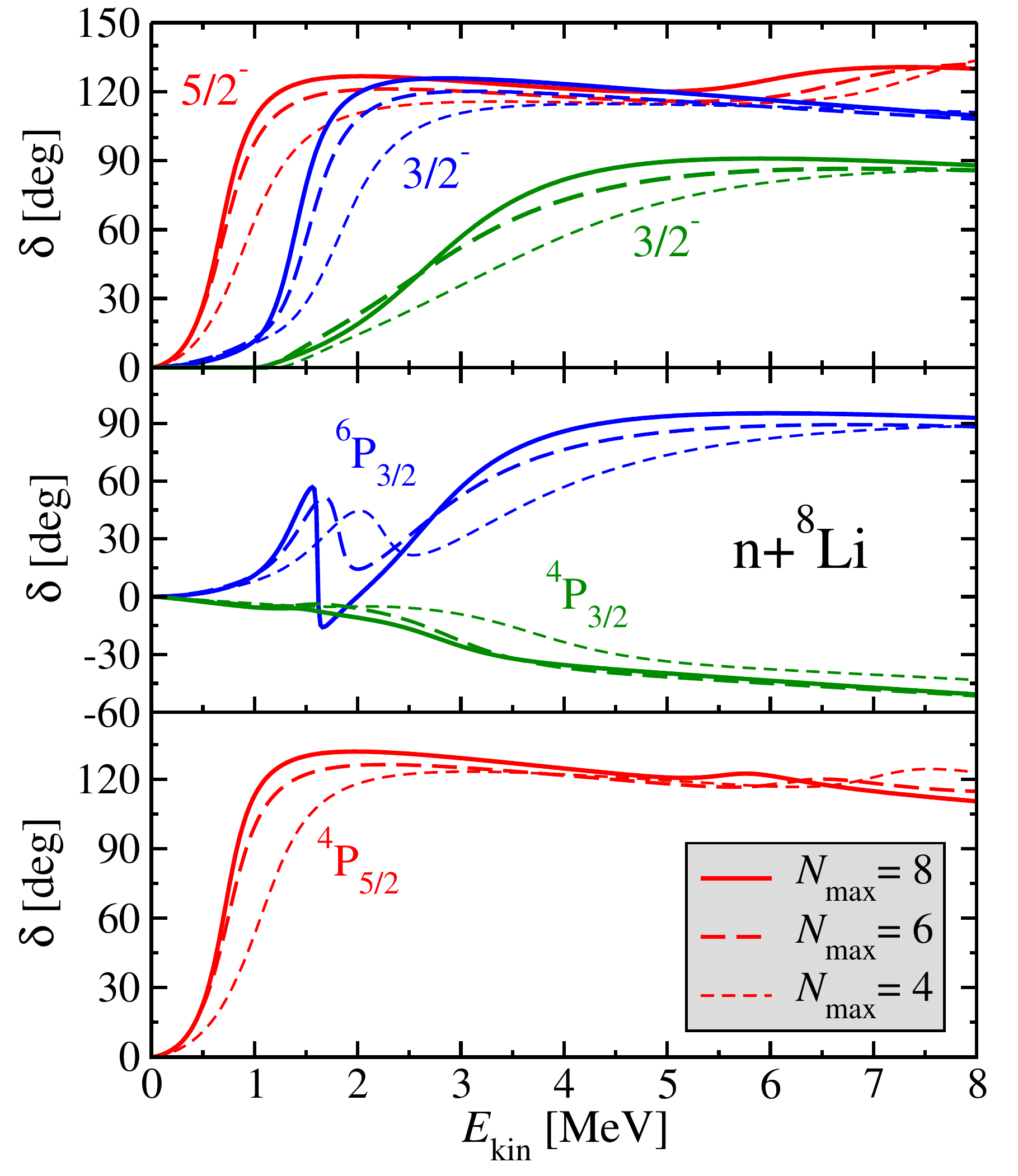}
\caption{\label{fig:eigenphase_conv} Dependence of the $^8$Li+$n$ eigenphase shifts (top panel) on the NCSMC basis size characterized by $N_{\rm max}$ for low-lying $\frac{5}{2}^-$ (red) and $\frac{3}{2}^-$ (blue and green) resonances of $\isotope[9]{Li}$. The same for selected $\frac{3}{2}^-$ (middle panel) and $\frac{5}{2}^-$ (bottom panel) $P$-wave phase shifts. The NN+3N(lnl) chiral interaction was used. $E_{\rm kin}$ is the kinetic energy of $^8$Li+$n$ in the center-of-mass frame.}
\end{figure}

Below the $^8$Li+$n$ energy of 4 MeV in the center of mass, we find three $P$-wave resonances corresponding to two $3/2^-$ and a $5/2^-$ state. Corresponding eigenphase shifts and selected partial wave phase shifts are shown in Figs.~\ref{fig:eigenphase_conv}. The convergence with respect to $N_{\rm max}$ is quite satisfactory, especially for the two sharper resonances. We note that the eigenphase shifts are obtained from the $S$-matrix eigenvalues while the partial wave phase shifts are obtained from diagonal matrix elements of the $S$-matrix. 

\begin{figure*}
\centering
\includegraphics[width=\textwidth]{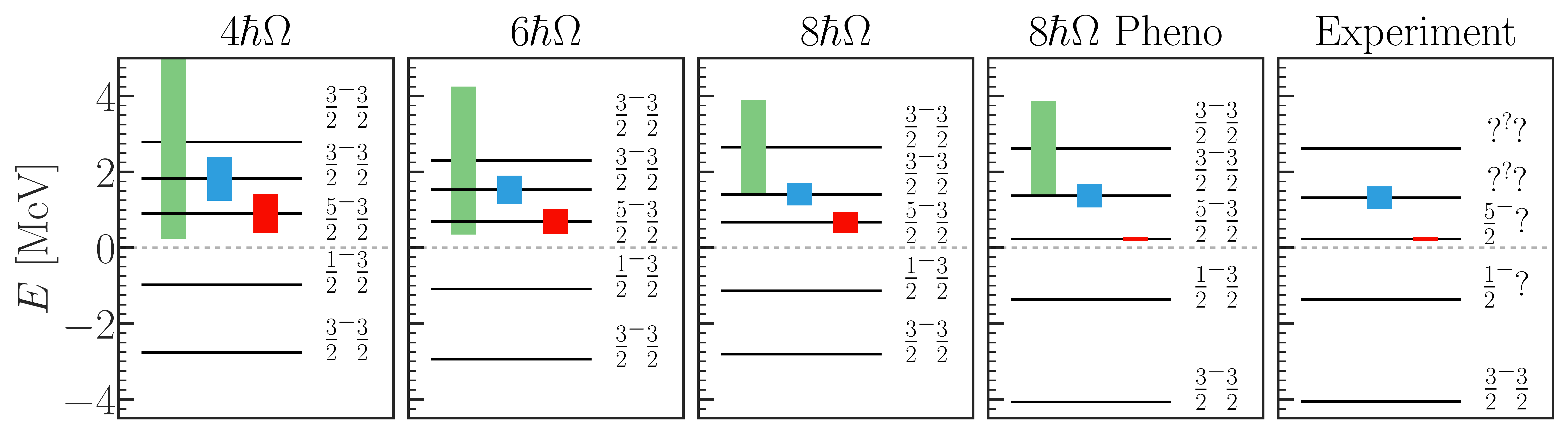}
\caption{\label{fig:scheme_all} 
Energies of $\isotope[9]{Li}$ bound states and low-lying resonances with respect to the $^8$Li+$n$ threshold. The leftmost three panels show NCSMC calculations at $N_{\rm max}{=} 4, 6$, and $8$. The fourth panel shows the NCSMC-pheno $N_{\rm max}{=} 8$ calculation. The NN+3N(lnl) chiral interaction was used. Coloured bars represent the widths of resonances. Experimental data in the rightmost panel are from Ref.~\cite{TILLEY2004155}. Question marks are used where data is unavailable.}
\end{figure*}

\begin{figure}
\centering
\includegraphics[width=0.47\textwidth]{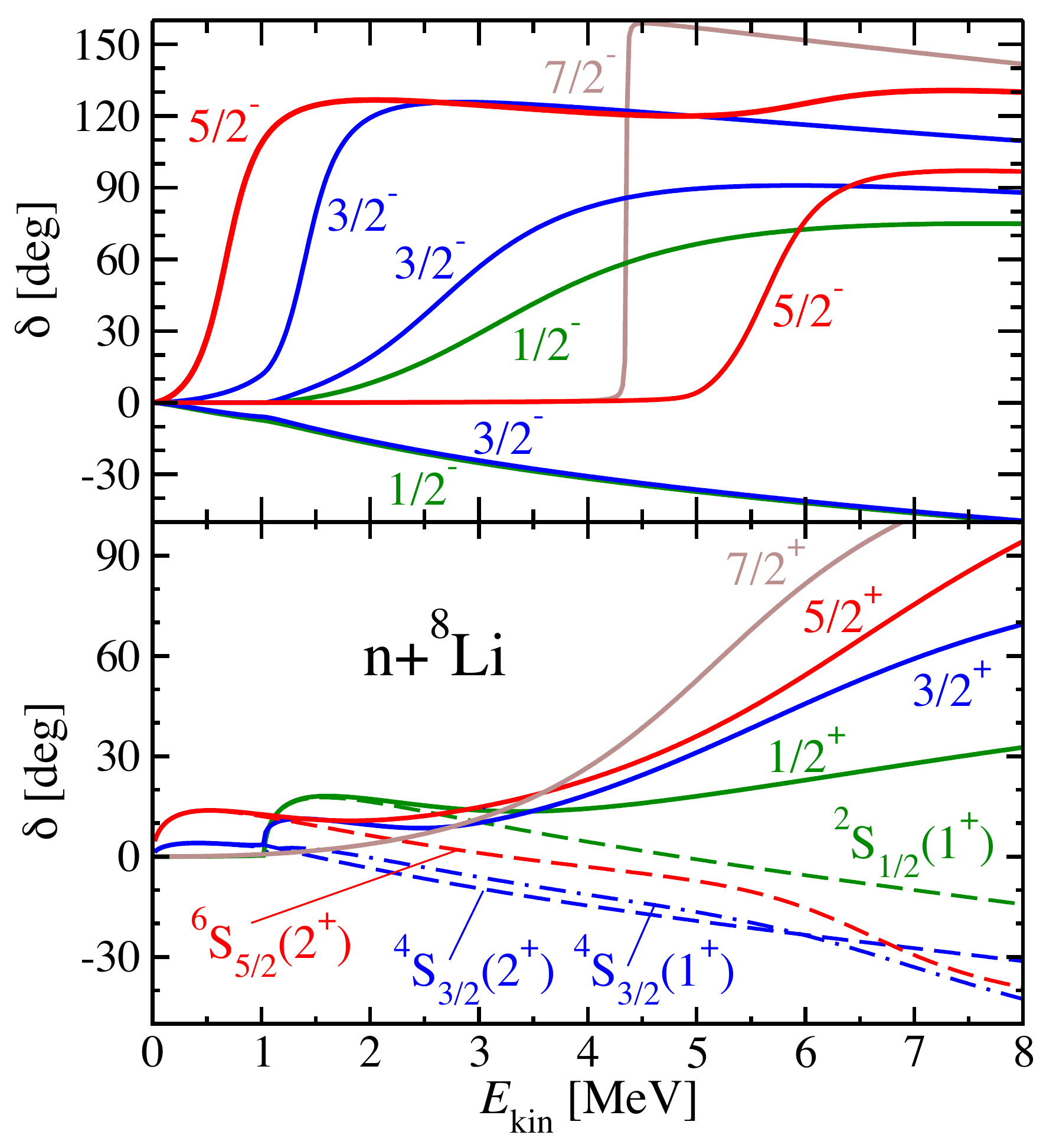}
\caption{\label{fig:eigenphase_8} $^8$Li+$n$ eigenphase shifts for selected negative-parity (top panel) and positive-parity (bottom panel) channels. Dashed lines in the bottom panel represent $S$-wave phase shifts. NCSMC calculations performed in $N_{\rm max}{=}8$ space with the NN+3N(lnl) chiral interaction.}
\end{figure}

In the leftmost three panels of Fig.~\ref{fig:scheme_all}, we show the bound-state energies, in addition to the energies and widths of the three resonances for the $N_{\rm max}{=}4,6,8$ model spaces. These are shown alongside available experimental data. The numerical values for the $N_{\rm max}{=}8$ space are then given in Table~\ref{tab:sepresen}. Selected eigenphase shifts and $S$-wave phase shifts obtained in the $N_{\rm max}{=}8$ space are presented in Fig.~\ref{fig:eigenphase_8}. It is clear that the calculated $5/2^-$ resonance is a good match to the experimentally known resonance at 4.296 MeV. We predict that the 5.38 MeV level is $3/2^-$. On the other hand, the experimentally very narrow 6.43 MeV level does not correspond to our calculated very broad second $3/2^-$. Rather, it presumably corresponds to the calculated $7/2^-$ state shown in the right panel of Fig.~\ref{fig:ncsd_spectra} and in the top panel of Fig.~\ref{fig:eigenphase_8} as an extremely narrow resonance. For a more realistic description of this state, we would most likely need to include the $3^+$ state of $^8$Li in the NCSMC cluster expansion~(\ref{eq_Li9_rgm_state})~\cite{Fortune2019}. The $^8$Li $3^+$ state that appears at 2.255 MeV in experiment (see the left panel of Fig.~\ref{fig:ncsd_spectra}) would obviously also impact other higher lying -- and in particular higher spin -- resonances, e.g., the $7/2^+$ and the second $5/2^+$, shown in Fig.~\ref{fig:eigenphase_8}. 

The decreasing $3/2^-$ and $1/2^-$ eigenphase shifts that start at $\delta{=}0^{\rm o}$ in Fig.~\ref{fig:eigenphase_8} correspond to the two bound states. On the other hand, all calculated $S$-wave phase shifts and their associated eigenphase shifts are rising at their respective thresholds, i.e., the corresponding scattering lengths are negative. In particular, for the $^6S_{5/2}(2^+)$ partial wave we find the scattering length of -0.44 fm while for $^4S_{3/2}(2^+)$ -0.13 fm. We note that a broad $5/2^+$ $T{=}3/2$ resonance in $^9$Be, an isospin analog of a resonance in $^9$Li, was very recently reported in Ref.~\cite{PhysRevC.102.014615}. It was found below the $T{=}3/2$ $5/2^-$ resonance, the isospin analog of the 4.296 MeV resonance in $^9$Li. 

Before proceeding with the calculation of the capture cross section, the NCSMC results were phenomenologically adjusted to reproduce experimental thresholds and positions of known resonances in an approach known as NCSMC-pheno ~\cite{DOHETERALY2016430,PhysRevLett.117.242501}. 
This step is necessary to obtain a quantitative evaluation of the capture cross section. The resulting evaluation embodies an advanced microscopic understanding of the underlying nuclear structure and reaction mechanism obtained from a chiral NN+3N Hamiltonian, but is no-longer a purely theoretical prediction.  
The phenomenological modifications are rather small and were accomplished first by adjusting the $^8$Li excitation energy of the $1^+$ state to its experimental value and, second, by fitting the $^9$Li NCSM input energies to reproduce the experimental $^9$Li energies in the NCSMC calculations. We performed the NCSMC-pheno calculations for the $N_{\rm max}{=}6$ and $N_{\rm max}{=}8$ model spaces. As seen in the left panel of Fig.~\ref{fig:ncsd_spectra}, our calculated excitation energy for the $^8$Li $1^+$ state is quite close to experiment. Consequently, it only needs a $-45$ keV adjustment in the $N_{\rm max}{=}8$ calculation. 
Next, we adjust the lowest NCSM $^9$Li eigenenergies in the $3/2^-$, $1/2^-$ and $5/2^-$ channels (used as input in the NCSMC calculation) to reproduce the experimental separation energies of the $3/2^-$ and $1/2^-$ bound states and the $5/2^-$ resonance centroid energy. As seen in the middle panel of Fig.~\ref{fig:scheme_all}, the NCSMC $1/2^-$ energy is already quite close to experiment, therefore a shift of -0.3 MeV in the lowest $1/2^-$ NCSM eigenvalue is sufficient to reproduce the separation energy. For the $3/2^-$ and $5/2^-$ channels, we need to modify the eigenvalues by about -1 MeV, i.e., 2.5\% of the calculated ground-state (g.s.) energy. 

\begin{figure}
\centering
\includegraphics[width=0.47\textwidth]{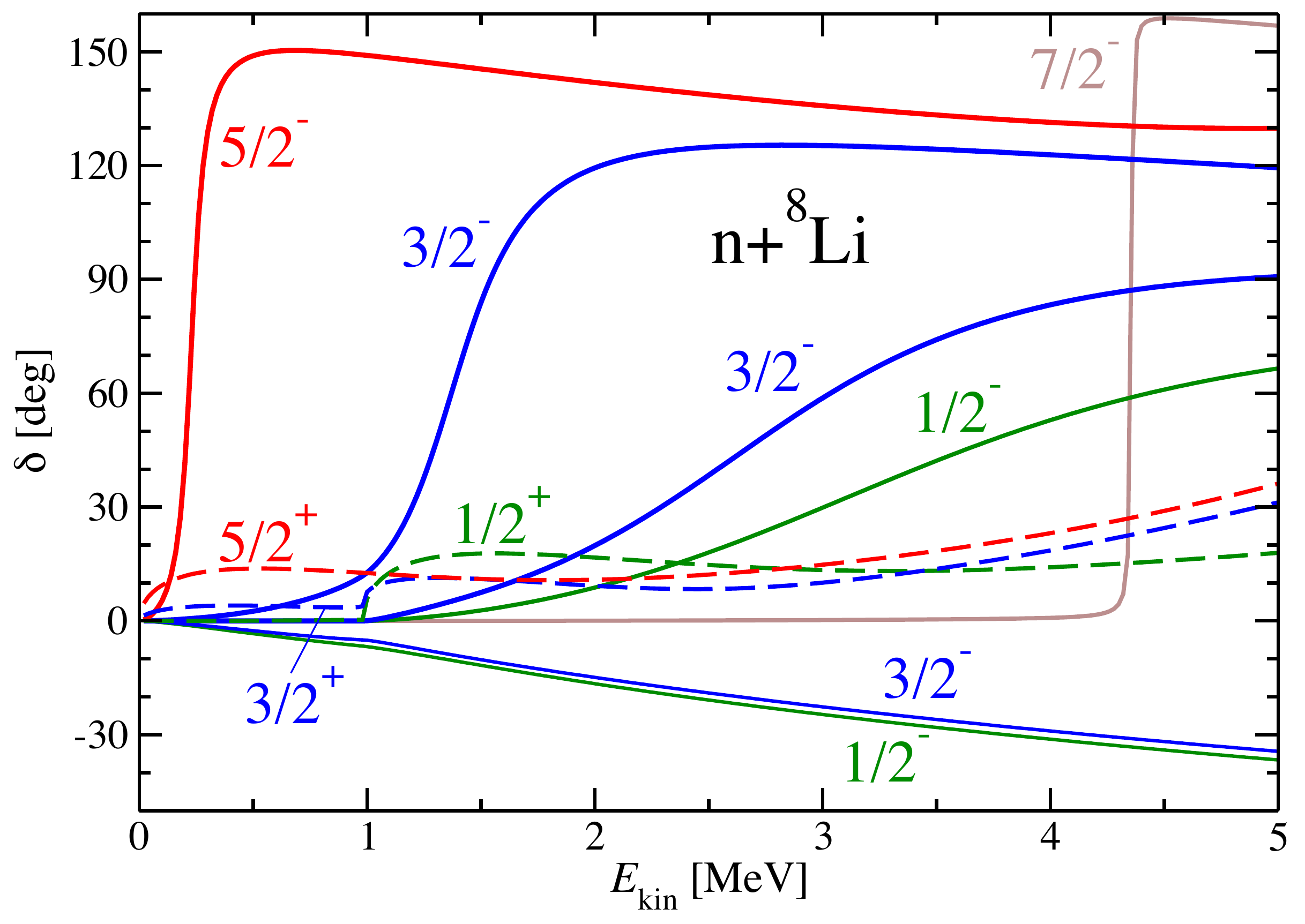}
\caption{\label{fig:eigenphase_8_pheno}  
$^8$Li+$n$ eigenphase shifts obtained from the $N_{\text{max}}=8$ NCSMC-pheno calculation for selected negative-parity (solid) and positive-parity (dashed) channels.}
\end{figure}

\begin{table}
\begin{center}
\begin{ruledtabular}
\begin{tabular}{ccccccc}
             & \multicolumn{2}{}{}NCSMC & \multicolumn{2}{}{}NCSMC-pheno & \multicolumn{2}{}{}Expt \\
 $J^\pi\; T$    &   $E$  & $\Gamma$ &   $E$  & $\Gamma$ &    $E$  & $\Gamma$ \\ 
 $3/2^-\; 3/2$  &  2.65  &  2.5(4)  &   2.62 &  2.5(4)  &    N/A  &    N/A  \\
 $3/2^-\; 3/2$  &  1.41  &  0.59    &   1.37 &  0.61    &   1.32$^{\rm a}$  & 0.60(10)\footnote{Experimental spin and parity assignment uncertain} \\
 $5/2^-\; 3/2$  &  0.67  &  0.56    &   0.23 &  0.11    &   0.23  & 0.10(3) \\
 $1/2^-\; 3/2$  & -1.14  &  -       &  -1.37 & -        &  -1.37  & - \\  
 $3/2^-\; 3/2$  & -2.81  &  -       &  -4.07 & -        &  -4.06  & - \\
\end{tabular}
\end{ruledtabular}
\caption{$^{9}$Li bound-state and resonance energies with respect to the $^8$Li${+}n$ threshold with the corresponding resonance widths. All values in MeV. NCSMC and NCSMC-pheno calculations have been performed with the NN+3N(lnl) chiral interaction in the $N_{\rm max}{=}8$ space. Experimental data are from Ref.~\cite{TILLEY2004155}.}
\label{tab:sepresen}
\end{center}
\end{table}

The resulting NCSMC-pheno bound-state energies, centroids and widths of the lowest three calculated resonances and selected eigenphase shifts for the $N_{\rm max}{=}8$ model space are presented in the fourth panel of Fig.~\ref{fig:scheme_all} and in Fig.~\ref{fig:eigenphase_8_pheno}, respectively. Due to the negligible adjustment of the $^8$Li $1^+$ energy, channels other than the $1/2^-, 3/2^-, 5/2^-$ are basically unmodified compared to the original NCSMC calculation. 

In Table~\ref{tab:sepresen}, we summarize bound-state energies, as well as centroid energies and widths of the lowest three calculated resonances obtained in the $N_{\rm max}{=}8$ NCSMC and NCSMC-pheno calculations. Within the table, these are compared to available experimental data. The resonance energies and width have been determined from the eigenphase shift derivatives as well as from an $S$-matrix analysis in the complex momentum space. The two methods agree very well for all the resonance energies and the widths of the two sharper resonances while they give a few hundred keV differences for the width of the broad $3/2^-$ resonance. This could be interpreted as a theoretical uncertainty, indicated in the table. We re-iterate that only the bound-state energies and the $5/2^-$ resonance energy were fitted in the NCSMC-pheno calculations. The widths of the resonances are predictions as well as the energies of the two $3/2^-$ states. Our calculations reproduce very well the experimental properties of the $5/2^-$ resonance, and the first calculated $3/2^-$ resonance matches the energy and width of the experimental 5.38 MeV state.

\begin{table}
\begin{center}
\begin{ruledtabular}
\begin{tabular}{ccccccc}
                  & ANC [fm$^{-1/2}$] & SF NCSM  & SF NCSMC-pheno \\
$^4P_{3/2} (2^+)$ &  1.026 &  0.64    &  0.59  \\
$^6P_{3/2} (2^+)$ &  0.995 &  0.41    &  0.41  \\
$^2P_{3/2} (1^+)$ & -1.009 &  0.39    &  0.37  \\
$^4P_{3/2} (1^+)$ & -0.663 &  0.11    &  0.11  \\
\end{tabular}
\end{ruledtabular}
\caption{$^9$Li $3/2^-$ g.s. asymptotic normalization coefficients (ANC) obtained in the NCSMC-pheno calculations and spectroscopic factors (SF) obtained in the NCSM and NCSMC-pheno. 
Calculations were performed in the $N_{\rm max}{=}8$ space. See Fig.~\ref{fig:clusterff} for other details.}
\label{tab:anc}
\end{center}
\end{table}

\begin{figure}
\centering
\includegraphics[width=0.5\textwidth]{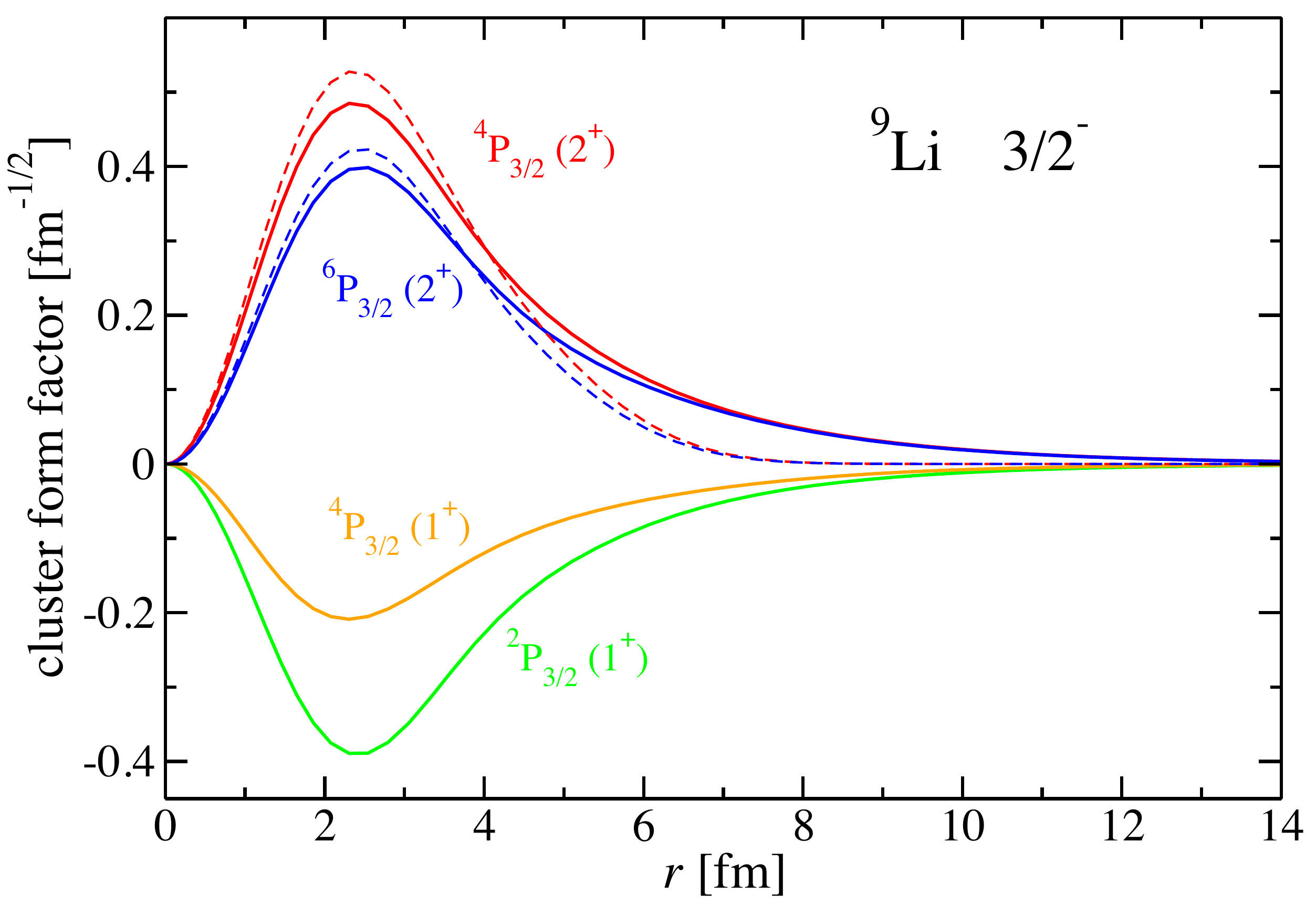}
\caption{\label{fig:clusterff} $^9$Li $3/2^-$ g.s. cluster form factors. Only $P$-wave components are shown. The full lines represent the $N_{\rm max}{=}8$ NCSMC-pheno calculations, the dashed lines (for the $^8$Li $2^+$ state channels only) are NCSM results. The coupling between the $^8$Li and neutron in the cluster state is given in Eq.~(\ref{eq_Li9_rgm_state}).}
\end{figure}

A realistic description of the structure of the $^9$Li ground state is essential for the description of the capture reaction. In Fig.~\ref{fig:clusterff}, we show the cluster form factors (overlap functions) for the $^9$Li $3/2^-$ g.s., defined by $r \bra{\Phi^{J^\pi T}_{\nu r, -\frac{3}{2}}} {\mathcal{A}}_\nu \ket{\Psi^{J^\pi T}_{A\texttt{=}9, -\frac{3}{2}}}$ with the states from Eqs. (\ref{ncsmc_wf_Li9}) and (\ref{eq_Li9_rgm_state}). The dashed lines represent the NCSM cluster form factors that serve as input to the NCSMC equations~\cite{,PhysRevC.87.034326,physcripnavratil}. While the NCSMC-pheno overlaps extend beyond $n$-$^8$Li separations of 10 fm, the NCSM ones are basically zero starting at 7 fm. By integrating the overlap functions squared, one obtains spectroscopic factors (SF), which we present in Table~\ref{tab:anc} together with the asymptotic normalization coefficients (ANCs). Although the NCSM and NCSMC-pheno cluster form factors differ, the spectroscopic factors are very similar. Still, we observe some reduction when continuum microscopic cluster states are included. The $^9$Li(g.s.)$\leftrightarrow ^8$Li(g.s.)+$n$ NCSMC-pheno total SF, 1.00, is in good agreement with the experimental value of 0.90(13) reported in Ref.~\cite{Wuosmaa2005}. Smaller SFs were reported in Refs.~\cite{Li2005} (0.68(14)),  \cite{Guimaraes2007} (0.62(13)), and~\cite{KANUNGO200826} (0.65(15)). Overall, our total NCSMC-pheno SFs, 1.00 ($^9$Li(g.s.)$\leftrightarrow ^8$Li($2^+$)+$n$) and 0.48 ($^9$Li(g.s.)$\leftrightarrow ^8$Li($1^+$)+$n$), are in excellent agreement with those obtained within the Variational Monte Carlo (VMC) method with Argonne and Illinois interactions,  0.97 and 0.46, respectively~\cite{PhysRevC.66.044310,VMCoverlaps}. Our calculated ANC values can be compared to the experimental determination of (ANC)$^2 {=}1.33(33)$ fm$^{-1}$ obtained from the angular distribution analysis of the $^8$Li($d,p$)$^9$Li$_{\rm gs}$ transfer reaction~\cite{GUO2005162}. A slightly smaller (ANC)$^2 {=}0.92(14)$ fm$^{-1}$ was reported in Ref.~\cite{2006isna.confE.108G} which is in excellent agreement with our calculations.

\subsection{$^8$Li($n,\gamma$)$^9$Li radiative capture}

We use a standard one-body electromagnetic current in the long wave length approximation taking into account E1, M1, and E2 multipolarities. In particular, the electric dipole operator can be cast in the form
\begin{equation}\label{eq:E1}
\hat{D}= e \sqrt{\frac{4\pi}{3}}\sum_{i=1}^A \frac{\tau_i^z}{2} r_i Y_{10}(\hat{r}_i) \;,
\end{equation}  
with $\tau_i^z$ and $\vec{r}_i=r_i\hat{r}_i$  representing the isospin third component and center of mass frame coordinate of the {\it i}th nucleon. This form of the E1 transition operator includes the leading effects of the meson-exchange currents through the Siegert$^,$s theorem. Two-body currents are expected to play a role for M1 transitions~\cite{Pastore:2015dho}. Since the capture proceeds dominantly by E1 radiation, we neglect the M1 two-body currents. As we utilize an SRG evolved chiral Hamiltonian, the electromagnetic transition operators should also be consistently SRG evolved. Such step is not taken in this work. 
In general, the SRG transformation is mostly driven by short range correlations in the NN interaction and its effect on long-range operators is rather small~\cite{Schuster2014,Schuster2015,Gysbers2019}. 

\begin{figure}
\centering
\includegraphics[width=0.49\textwidth]{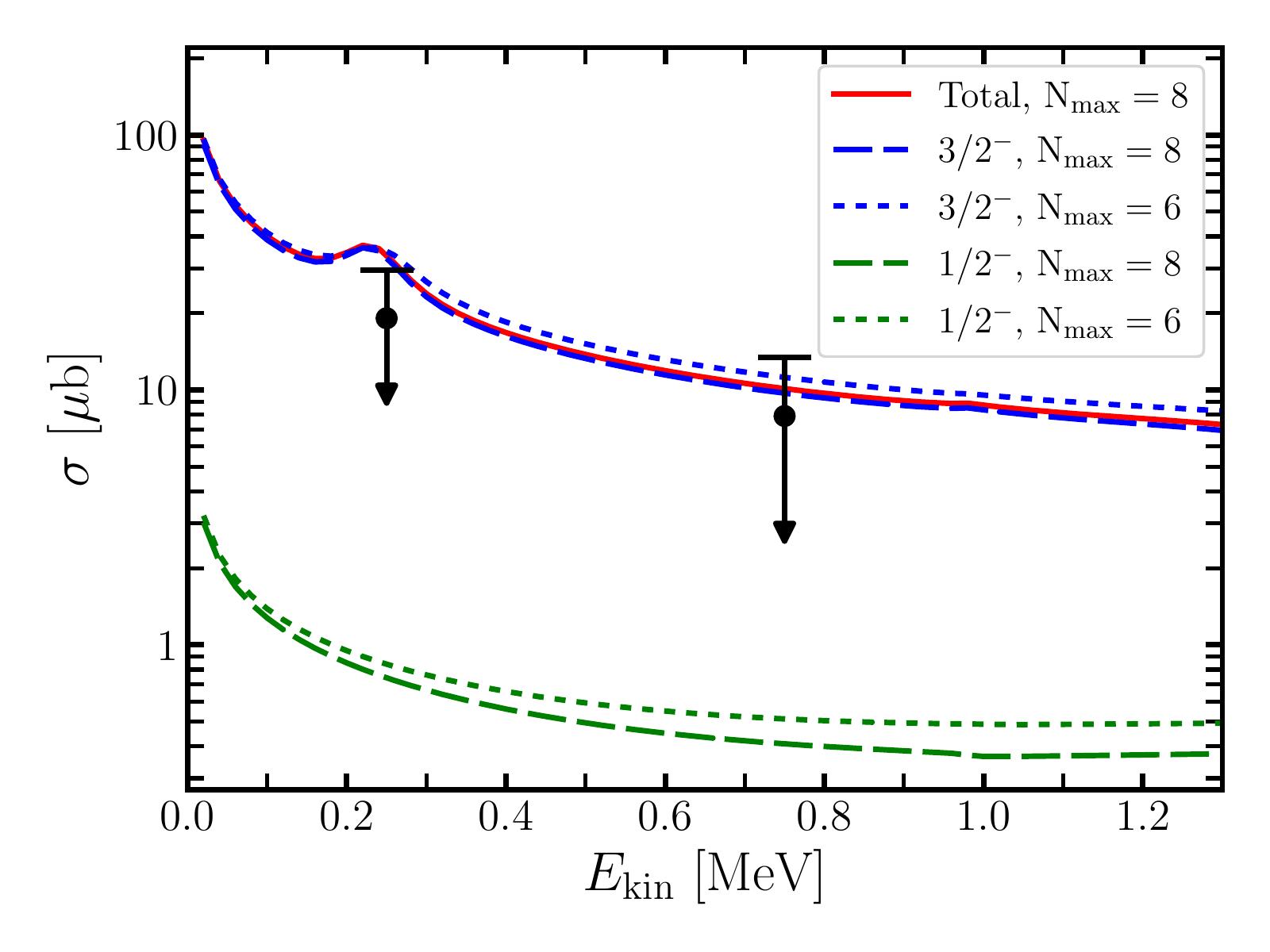}
\caption{\label{fig:cross_section} $\isotope[8]{Li}(n, \gamma)\isotope[9]{Li}$ capture cross section obtained in the NCSMC-pheno calculations. We compare $N_{\rm max}{=}6$ (dotted lines), $N_{\rm max}{=}8$ (dashed lines), the total $N_{\rm max}{=}8$ cross-section (solid line), and experimental limits from Ref.~\cite{Zecher1998} (black points). Cross-section contributions from the ground state are shown in blue, contributions from the first excited state are in green.}
\end{figure}

Our calculated $\isotope[8]{Li}(n, \gamma)\isotope[9]{Li}$ capture cross section is presented in Fig.~\ref{fig:cross_section}. We compare NCSMC-pheno results obtained in the $N_{\rm max}{=}8$ and $N_{\rm max}{=}6$ spaces. Overall, we find a good stability of the calculations. By increasing the model space, the cross section gets reduced slightly and the difference can serve as an estimate of the uncertainty. The capture to the $^9$Li ground state dominates the total cross section. The excited state contribution is suppressed by more than an order of magnitude. In the low-energy region displayed in Fig.~\ref{fig:cross_section}, the non-resonant E1 capture is the leading contribution. The E2/M1 capture enhanced by the $5/2^-$ resonance is visible as a bump around 0.23 MeV.

Our calculated cross section is on the higher side but still within the limits derived from the 1998 NSCL Coulomb dissociation experiment~\cite{Zecher1998} shown in Fig.~\ref{fig:cross_section} by black points and vertical lines. These limits should be compared to the E1 contribution to the capture to the ground state.   

\begin{figure}
\centering
\includegraphics[width=0.49\textwidth]{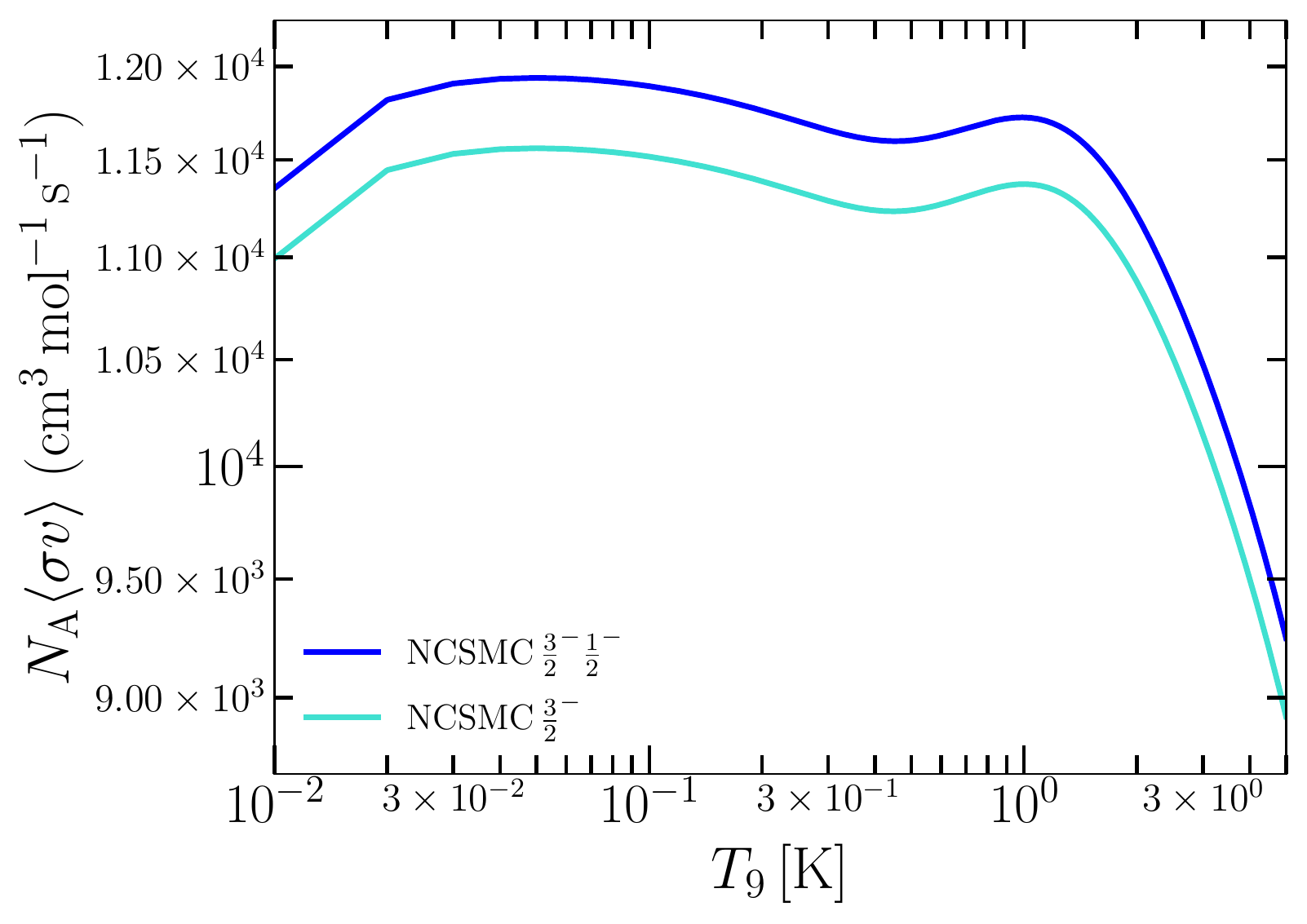}
\caption{\label{fig:rate} $\isotope[8]{Li}(n, \gamma)\isotope[9]{Li}$ reaction rate obtained in the $N_{\rm max}{=}8$ NCSMC-pheno calculations. The upper line shows the total reaction rate, and the lower line shows the ground-state contribution.}
\end{figure}

The $\isotope[8]{Li}(n, \gamma)\isotope[9]{Li}$ reaction rate obtained from our total capture cross section is shown in Fig.~\ref{fig:rate}. In addition, we present the contribution of the capture to the ground state to the overall reaction rate. Our results are smaller by a factor of 4 and 2 compared to values reported in Refs.~\cite{Malaney1989} and \cite{MAO1991568}, respectively. However, they are higher by a factor of 2 compared to the recent potential cluster model calculations from Ref.~\cite{Dubovichenko_2016}. One of the reasons for the smaller reaction rate obtained in the latter calculations is the lower value of the spectroscopic factor used as input for the potential cluster model calculations compared to the spectroscopic factor obtained as an output of our many-body calculations.

\section{Conclusions}
\label{sec_conclusions}

We applied the {\it ab initio} NCSMC to study properties of $^9$Li bound states and low-lying resonances, and calculated the $^8$Li($n,\gamma$)$^9$Li cross section. Chiral nucleon-nucleon and three-nucleon interactions from Refs.~\cite{Entem2003} and \cite{PhysRevC.101.014318} served as input for our calculations, though for the purpose of predicting the capture cross section we  adjusted the thresholds and the position of the lowest resonance to their experimental values.

Our calculations reproduce experimentally known bound states as well as the lowest $5/2^-$ resonance of $^9$Li. We predict the 5.38 MeV resonance to be a $3/2^-$ state. In addition to the very narrow $7/2^-$ resonance, corresponding most likely to the experimental 6.43 MeV state, we find several other broad low-lying resonances. In particular, at ~2.6 MeV above the $^8$Li+$n$ threshold we find a broad $3/2^-$ resonance with the width of 2.5 MeV. The description of the $7/2^-$ resonance and of the higher lying $7/2^+$ and $5/2^\pm$ resonances can be improved by including the $^8$Li $3^+$ state in the NCSMC trial wave function (Eqs.~(\ref{ncsmc_wf_Li9}), (\ref{eq_Li9_rgm_state})). We plan to perform such calculations in the future.

Our calculated $^8$Li($n,\gamma$)$^9$Li capture cross section is on the higher side but within the limits derived from the 1998 NSCL Coulomb dissociation experiment. It is dominated by the direct E1 capture to the ground state with a resonant contribution around 0.23 MeV due to E2/M1 radiation enhanced by the $5/2^-$ resonance. 
 
The reaction rate obtained from our calculated capture cross section is lower than early evaluations. However, it is higher by about a factor of two compared to recent potential cluster model calculations. Our results indicate that the $^8$Li($n,\gamma$)$^9$Li reaction might play a more important astrophysical role than recently considered. 

Results presented in this paper demonstrate current capabilities of the NCSMC. With high-precision chiral NN+3N interactions as the input, we are able to predict with confidence properties of light nuclei even with a large neutron excess. NCSMC calculations of several other radiative capture reactions important for astrophysics including $^{7}$Be($p,\gamma$)$^{8}$B, $^{11}$C($p,\gamma$)$^{12}$N, and $^{14}$C($n,\gamma$)$^{15}$C are under way. 
In the future we plan on quantifying the uncertainty related to the convergence of the chiral expansion (before and after the application of phenomenological corrections) by extending the methodology of Ref.~\cite{Kravvaris2020} to capture cross sections.

\acknowledgments

We thank Peter Gysbers and Mack Atkinson for useful discussions. This work was supported by the NSERC Grant No. SAPIN-2016-00033 and by the U.S. Department of Energy, Office of Science, Office of Nuclear Physics, under Work Proposals No. SCW0498. TRIUMF receives federal funding via a contribution agreement with the National Research Council of Canada, and TRIUMF's student program is funded in part by RBC. This work was prepared in part by LLNL under Contract No. DE-AC52-07NA27344. Computing support came from an INCITE Award on the Summit supercomputer of the Oak Ridge Leadership Computing Facility (OLCF) at ORNL, from Westgrid and Compute Canada.

\FloatBarrier

%

\end{document}